# High-*Q* silk fibroin whispering gallery microresonator

**Linhua Xu,[1,2] Xuefeng Jiang,[1] Guangming Zhao,[1] Ding Ma,[3] Hu Tao,[4] Zhiwen Liu,[3] Fiorenzo G. Omenetto,[4] and Lan Yang[1,2,*]**

[1]*Electrical and Systems Engineering Department, Washington University in St. Louis, St. Louis, MO 63130, USA*
[2]*Institute of Materials Science and Engineering, Washington University in St. Louis, St. Louis, MO 63130, USA*
[3]*Department of Electrical Engineering, The Pennsylvania State University, University Park, PA 16802, USA*
[4]*Department of Biomedical Engineering, Tufts University, Medford, MA 02155, USA*
*\* yang@seas.wustl.edu*

**Abstract:** We have experimentally demonstrated an on-chip all-silk fibroin whispering gallery mode microresonator by using a simple molding and solution-casting technique. The quality factors of the fabricated silk protein microresonators are up to $10^5$. A high-sensitivity thermal sensor was realized in this silk fibroin microtoroid with sensitivity of 1.17 nm/K, 8 times higher than previous WGM resonator based thermal sensors. This opens the way to fabricate biodegradable and biocompatible protein based microresonators on a flexible chip for biophotonics applications.

**OCIS codes:** (140.3945) Microcavities; (140.4780) Optical resonators; (160.1435) Biomaterials.

## References and links

1. K. J. Vahala, "Optical microcavities," Nature **424**(6950), 839–846 (2003).
2. F. Vollmer and L. Yang, "Label-free detection with high-*Q* microcavities: a review of biosensing mechanisms for integrated devices," Nanophotonics, **1**(3-4), 267-291 (2012).
3. M. R. Foreman, J. D. Swaim, and F. Vollmer, "Whispering gallery mode sensors," Advances in Optics and Photonics, **7**(2), 168-240 (2015).
4. X. Fan, I. M. White, S. I. Shopova, H. Zhu, J. D. Suter, and Y. Sun, "Sensitive optical biosensors for unlabeled targets: A review," Analytica Chimica Acta, **620**(1), 8-26 (2008).
5. F. Vollmer and S. Arnold, "Whispering-gallery-mode biosensing: label-free detection down to single molecules," Nature Methods **5**(7), 591–596 (2008).
6. M. D. Baaske, M. R. Foreman, and F. Vollmer, "Single-molecule nucleic acid interactions monitored on a label-free microcavity biosensor platform," Nature Nanotech. **9**, 933-939 (2014).
7. T. Lu, H. Lee, T. Chen, S. Herchakb, J.-H. Kim, S. E. Frasera, R. C. Flagand, and K. Vahala, "High sensitivity nanoparticle detection using optical microcavities," Proc. Natl. Acad. Sci. U.S.A. **108**, 5976–5979 (2011).
8. Y.-F. Xiao, V. Gaddam, and L. Yang, "Coupled optical microcavities: an enhanced refractometric sensing configuration," Opt. Express **16**, 12538-12543 (2008).
9. J. Zhu, S. K. Ozdemir, Y. F. Xiao, L. Li, L. He, D. R. Chen, and L. Yang, "On-chip single nanoparticle detection and sizing by mode splitting in an ultrahigh-*Q* microresonator," Nature Photonics **4**(1), 46–49 (2010).
10. L. He, S. K. Ozdemir, J. Zhu, W. Kim, and L. Yang, "Detecting single viruses and nanoparticles using whispering gallery microlasers," Nat. Nanotechnol. **6**, 428–432 (2011).
11. L. Shao, X.-F. Jiang, X.-C. Yu, B.-B. Li, W. R. Clements, F. Vollmer, W. Wang, Y.-F. Xiao, and Q. Gong, "Detection of single nanoparticles and lentiviruses using microcavity resonance broadening," Adv. Mater. **25**(39), 5616–5620 (2013).
12. B. D. Lawrence, M. Cronin-Golomb, I. Georgakoudi, D. L. Kaplan, and F. G. Omenetto, "Bioactive silk protein biomaterial systems for optical devices," Biomacromolecules **9**(4), 1214–1220 (2008).
13. Y. L. Sun, Z. S. Hou, S. M. Sun, B. Y. Zheng, J. F. Ku, W. F. Dong, Q. D. Chen and H. B. Sun, "Protein-based three-dimensional whispering-gallery-mode micro-lasers with stimulus-responsiveness", Sci. Rep. **5**, 12852 ( 2015).
14. H. Tao, D. L. Kaplan, and F. G. Omenetto, "Silk Materials - A Road to Sustainable High Technology," Adv. Mater. **24**, 2824–2837 (2012).
15. S. T. Parker, P. Domachuk, J. Amsden, J. Bressner, J. A. Lewis, D. L. Kaplan, and F. G. Omenetto, "Biocompatible Silk Printed Optical Waveguides," Adv. Mater. **21**, 2411–2415 (2009)
16. F. G. Omenetto and D. L. Kaplan, "New opportunities for an ancient material," Science **329**, 528–31 (2010).

17. J. MacLeod and F. Rosei, "PHOTONIC CRYSTALS Sustainable sensors from silk," Nat. Mater. **12**, 98–100 (2013).
18. D. N. Rockwood, R. C. Preda, T. Yücel, X. Wang, M. L. Lovett, and D. L. Kaplan, "Materials fabrication from Bombyx mori silk fibroin," Nat. Protoc. **6**, 1612–31 (2011).
19. H. Tao, J. M. Kainerstorfer, S. M. Siebert, E. M. Pritchard, A. Sassaroli, B. J. B. Panilaitis, M. A. Brenckle, J. J. Amsden, J. Levitt, S. Fantini, D. L. Kaplan, and F. G. Omenetto, "Implantable, multifunctional, bioresorbable optics," Proc. Nat. Acad. Sci. USA **109**, 19584-19589 (2012).
20. F. G. Omenetto and D. L. Kaplan, "A new route for silk," Nat. Photon. **2**, 641-643 (2008).
21. B.-B. Li, Q.-Y. Wang, Y.-F. Xiao, X.-F. Jiang, Y. Li, L. Xiao, and Q. Gong, "On chip, high-sensitivity thermal sensor based on high-$Q$ polydimethylsiloxane-coated microresonator," Appl. Phys. Lett. **96**(25), 251109 (2010).
22. C.-H. Dong, L. He, Y.-F. Xiao, V. R. Gaddam, S. K. Ozdemir, Z.-F. Han, G.-C. Guo, and L. Yang, "Fabrication of high-$Q$ polydimethylsiloxane optical microspheres for thermal sensing," Appl. Phys. Lett. **94**, 231119 (2009).
23. Y.-Z. Yan, C.-L. Zou, S.-B. Yan, F.-W. Sun, Z. Ji, J. Liu, Y.-G. Zhang, L. Wang, C.-Y. Xue, and W.-D. Zhang, "Packaged silica microsphere-taper coupling system for robust thermal sensing application," Opt. Express **19**, 5753–5759 (2011).
24. X. Xu, X. Jiang, G. Zhao, and L. Yang, "Phone-sized whispering-gallery microresonator sensing system," arXiv:1607.04651.
25. D. Saravanan, "Spider Silk - Structure, Properties and Spinning," J. Text. Apparel, Technol. Manag. **5**, 1–20 (2006).
26. D. K. Armani, T. J. Kippenberg, S. M. Spillane, and K. J. Vahala, "Ultra-high-$Q$ toroid microcavity on a chip," Nature, **421**(6926), 925-928 (2003).
27. A. L. Martin, D. K. Armani, L. Yang, and K. J. Vahala, "Replica-molded high-$Q$ polymer microresonators," Opt. Lett. **29**, 533-535 (2004).
28. B. Min, L. Yang, and K. Vahala, "Perturbative analytic theory of an ultrahigh-$Q$ toroidal microcavity," Phys. Rev. A **76**, 013823 (2007).
29. M. L. Gorodetsky, A. A. Savchenkov, and V. S. Ilchenko, "Ultimate $Q$ of optical microsphere resonators," Opt. Lett. **21**(7), 453-455 (1996).
30. M. K. Gupta, S. Singamaneni, M. McConney, L. F. Drummy, R. R. Naik, and V. V Tsukruk, "A facile fabrication strategy for patterning protein chain conformation in silk materials," Adv. Mater. **22**, 115–9 (2010).
31. A. Motta, L. Fambri, and C. Migliaresi, "Regenerated silk fibroin films: Thermal and dynamic mechanical analysis," Macromol. Chem. Phys. **203**, 1658–1665 (2002).
32. T. Carmon, L. Yang, and K. J. Vahala, "Dynamical thermal behavior and thermal self-stability of microcavities," Opt. Express **12**, 4742-4750 (2004).
33. L. He, Y.-F. Xiao, C. Dong, J. Zhu, V. Gaddam, L. Yang, "Compensation of thermal refraction effect in high-$Q$ toroidal microresonator by polydimethylsiloxane coating," Appl. Phys. Lett. **93**, 201102 (2008).

## 1. Introduction

Whispering gallery mode (WGM) microresonators, featuring ultrahigh quality ($Q$) factors and small mode volumes, significantly enhance light-matter interactions [1], and thus become an excellent platform for bio/chemical sensors [2-4] by monitoring the resonant frequency shift [5-8] or the scattering-induced mode splitting [9,10]/broadening [11]. So far, optical WGM resonator based biosensors are mainly fabricated in silica for all types of optical resonators [2], which limits the biosening applications in vivo because it may not only cause adverse physiologically reactions or immunological rejection but can also be difficult to be removed from the body. In fact, biodegradation and biocompatibility are two of the most important factors for in-vivo biosensing [12]. In this case, a WGM microresonator made from biomaterial possessing both biodegradation and biocompatibility will be an excellent candidate for in-vivo biosensing [13]. Silk fibroin originally from Bombyx mori (*i.e.*, silkworm), has desired properties for biosensing applications, including non-toxicity, biodegradation, biocompatibility, and high transparency in visible light band (>95 %) [14-20], making it a suitable material alternative for WGM resonators.

In this paper, we report the fabrication and characterization of all silk fibroin microtoroidal resonators with $Q$ factor on the order of $10^5$, dominated by material absorption loss. Furthermore, we have demonstrated thermal sensing based on silk fibroin microtoroid with the sensitivity as high as 1.17 nm/K, which is about 8 times higher than the previous WGM resonator based thermal sensors [21-24]. The high sensitivity of the silk resonator based thermal sensor originates from the large thermal expansion coefficient [25], which is three order of magnitude larger than that of silica.

## 2. Fabrication of the silk fibroin toroids

The fabrication process of silk fibroin microtoroidal resonators is illustrated in Fig. 1(a). Specifically, regenerated silk water solution was extracted from *Bombyx mori* silkworm cocoons as described in Ref. [18]. First, silk cocoons were cut into pieces and boiled in 0.02M $Na_2CO_3$ solution for 30 mins. Subsequently, the silk fibroin was dried overnight in a fume hood. Then the dried silk was dissolved in 9.3 M LiBr solution for 4 h at 60 °C, and the solution was dialyzed against ultra-pure water at room temperature for 48 h. Next, 6 wt% silk solution was obtained by removing the small impurities with centrifugation. Meanwhile, Ultrahigh $Q$ $SiO_2$ microtoroids were fabricated via standard processes described in Ref. [26], and then used to generate molds for the silk resonators [27]. As typical molding materials of negative phase mold, polydimethylsiloxane (PDMS) was poured onto microtoroids after being silanized with trichloromethylsixane, and then cured overnight at room temperature. Next, the resulting PDMS mold was filled with regenerated silk fibroin solution to form silk based microresonators after being released from microtoroid master. Afterwards, the sample was placed inside the fume hood overnight at room temperature to dry silk solution. Finally, the silk resonators were peeled away from the mold. Side-view and zoom-in top-view scanning electron microscope (SEM) images of a silk fibroin microtoroid are shown in Fig. 1(b), where an ultra-smooth surface can be seen.

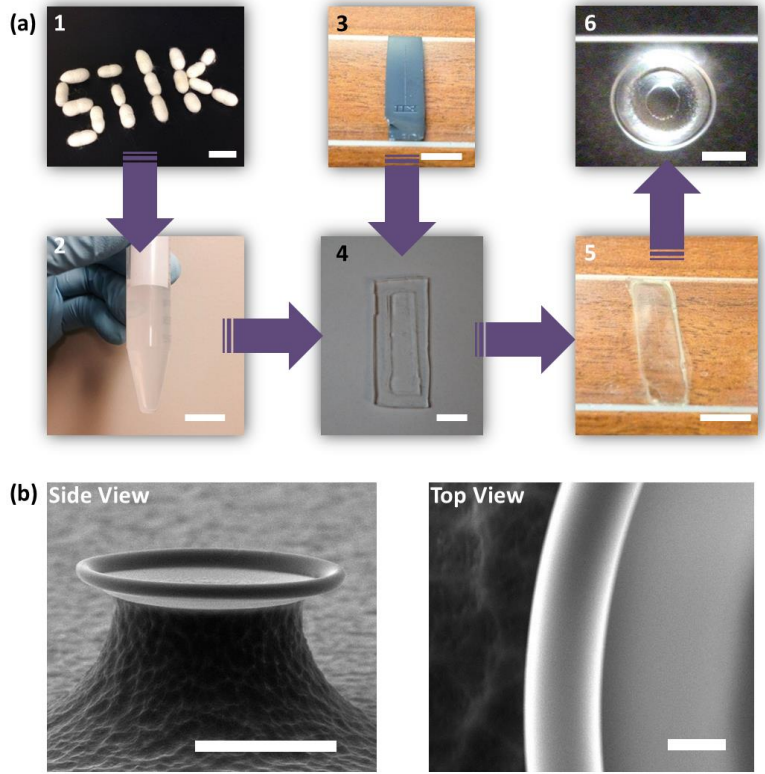

Fig. 1 (a) Flow diagram illustrating the fabrication process of the silk toroids. 1) Bombyx mori Silkworm cocoons. 2) Regenerated silk protein solution was extracted from silk cocoon. 3) On-chip microtoroid resonator array was obtained from state-of-art technique. 4) A negative phase PMDS mold mask was made from ultra-high-$Q$ microtoroids on a silicon chip. 5) Transparent film with silk fibroin microresonator was obtained by dry-casting silk solution in the PDMS mold. 6) Optical image of a silk microtoroid resonator with diameter of 80 μm. Scale bar: (1) 5cm, (2) 2cm, (3)-(5) 5mm, and (6) 40μm. (b) Side-view and inset top-view scanning electron microscopy (SEM) image of a silk fibroin microtoroid with diameter of 80 μm. Scale bar: 50 μm (sideview) and 5 μm (topview).

## 3. Characterization of optical properties of silk microtoroids

To characterize the optical properties of the silk fibroin microtoroids, a tunable diode laser in 1450 nm wavelength band (Newfoucs TLB 6327) was used to excite WGMs in a silk fibroin microtoroid. A fiber taper with diameter of around 1.3 μm was utilized to couple the laser light into and out of the microresonators, as shown in Fig. 2(a). The transmitted light was then detected by a photoreceiver (Newfocus 1811), and finally monitored by an oscilloscope (Tektronix TDS 3014B). A typical transmission spectrum of a silk fibroin microtoroid with diameter about 100 μm is shown in Fig. 2(b), where a $Q$ factor of $0.9 \times 10^5$ was found by Lorenzian fitting of the spectrum lineshape (red curve) in 1450 nm wavelength band. Remarkably, The free spectral range (FSR ~ 4.06 nm) of the resonator, as shown in Fig. 2(c), can be observed by scanning the laser wavelength over a large range (1431 nm – 1439 nm), which agrees well with the measured resonator diameter (d ~ 104 μm as SEM measurement).

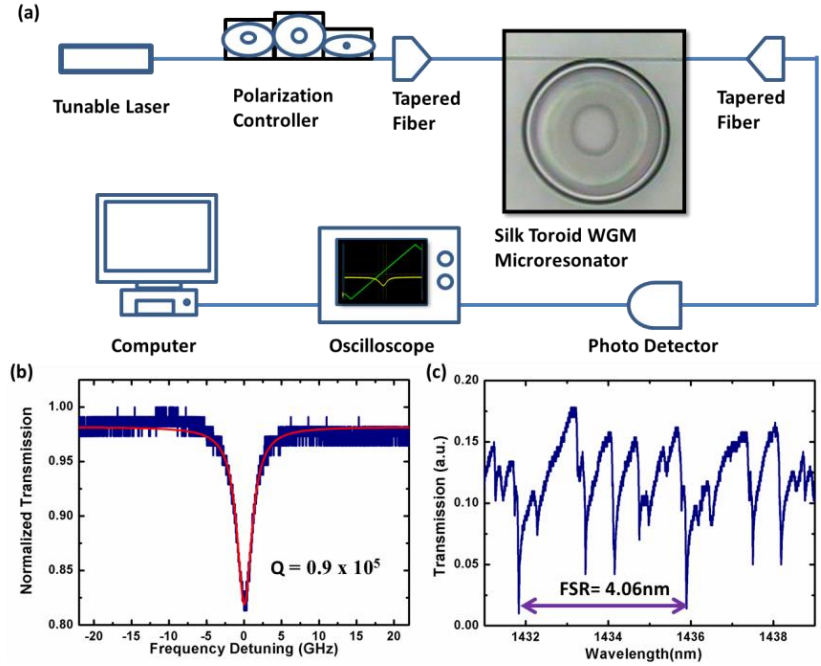

Fig. 2. (a) Illustration of the setup for testing and measuring the transmission spectrum of the silk fibroin microresonators. (b) Normalized transmission spectrum (blue curve) and the corresponding Lorentzian fitting (red curve) of a WGM in the silk fibroin microtoroid. (c) Wide-range transmission spectrum showing the free spectral range of the silk fibroin microtoroid.

Furthermore, we have also measured the $Q$ factors in different wavelength bands by utilizing several different tunable lasers and fiber tapers, including 670 nm, 770 nm, 980 nm and 1450 nm bands. Remarkably, the $Q$ factor in all these wavelength bands are on the order of $10^5$, specifically, with $Q$ factor of $2.3 \times 10^5$ in 670 nm band, $1.9 \times 10^5$ in 770 nm band, $2.4 \times 10^5$ in 970 nm band, and $0.9 \times 10^5$ in 1450 nm band. In general, the intrinsic $Q$ factor of a WGM microresonator is dominated by three kinds of losses: radiation loss, scattering loss and material absoption loss. The total $Q$ factor can be expressed as $Q_{total}^{-1} = Q_{rad}^{-1} + Q_{scat}^{-1} + Q_{material}^{-1}$. The radiation loss dominated $Q_{rad}$ is determined by the probe laser wavelength (670 nm - 1450 nm), the cavity diameter (100 μm), and the refractive index of the cavity material (1.55), and thus $Q_{rad} > 10^{10}$ can be easily obtained in the current case [28]. On the other hand, the scattering loss $Q_{scat}$ is dominated by the the average roughness ($R_a$) of the resonator exterior surface, which is about 1.849 nm for the silk toroid, confirmed by an atomic force microscopy

(AFM) image (inset of Fig. 3), implying a low scattering loss in silk microtoroid resonator; thus $Q_{scat} > 10^7$ is expected [29]. It should be noted that the long-range fluctuation of the toroid exterior surface is attributed to the curved surface of the toroidal structure [26]. Ultimately, silk's material absorption loss is the main factor limiting the $Q$ factor of the silk fibroin microtoroids in all the probe wavelength bands.

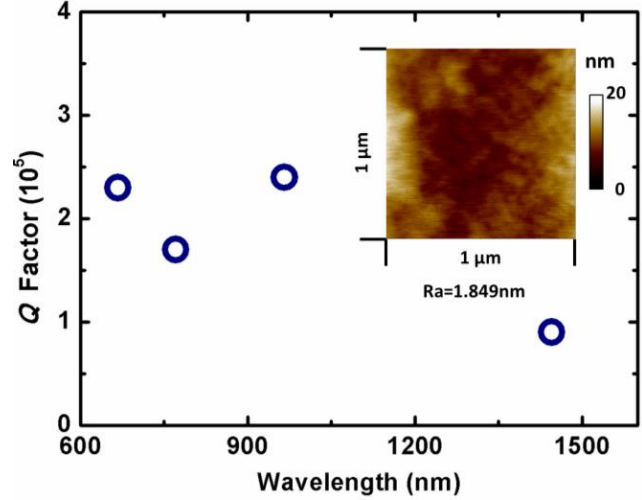

Fig. 3. Experimental $Q$ factor of silk microtoroidal resonator in different wavelength bands. Inset: Atomic force microscopy (AFM) measurement of toroidal exterior surface, with scanning area: 1 μm × 1 μm and $R_a$ = 1.264 nm, showing the surface smoothness.

## 4. Thermal sensor based on silk fibroin microtoroids

We have performed a thermal sensing experiment based on the fabricated silk fibroin microtoroids, as shown in Fig. 4. In the experiment, we placed a thermoelectric cooler (TEC) under the silk fibroin microtoroids to quantitatively change the local temperature by adjusting the driving current, and then recorded the resonant wavelength changes of a specific WGM at 1427.2 nm. A total wavelength shift of 3.98 nm as a result of 6.7 °C temperature change was observed with the sensitivity of -0.61 nm/K. In the experiment, silk fibroin was processed from water solution, thus silk microtoroid was dominated by *Silk I* (random coil and α-helix conformation) [30]. When well-oriented β-sheet crystalline conformation is dominant in the device, then the structure is in a so-called.*Silk II* condormation, which is the general state of natural silk fibers. Since *Silk II* has a more crystalline regine, it have a more sensitive thermal response [31]. Here, we used methanol to effectively crosslink the device and render it water-insoluble, thus enabling the transition from a *Silk I* to *Silk II* structure by inducing more H-binding within the material. Specifically, methanol vapor was generated in a glass dish on a hot plate at 40 °C. Then the silk toroidal resonators were suspended above the glass dish and exposed to methanol vapor for 4 minutes. The whole silk film was dried completely inside a fume hood at room temperature for 20 minutes after exposure was finished. The resulting device has a thermal response with a measured sensitivity of -1.17 nm/K (circle points in Fig. 4), which is about 8 times what was previously reported in WGM thermal sensor [21-24].

The resonant wavelength shift of WGM versus temperature can be theoretically analyzed as follow [32,33],

$$\frac{d\lambda}{dT} = \lambda_r \left( \frac{1}{n_{eff}} \frac{dn_{eff}}{dT} + \frac{1}{D} \frac{dD}{dT} \right) \quad (1)$$

where $\lambda_r$ is the resonant wavelength without temperature change, $n_{eff}$ and $D$ are the effective refractive index and the principal diameter of the silk fibroin toroid, respectively. The thermal expansion coefficient ($\alpha = dD/(DdT)$) of natural silk is $-10.6 \times 10^{-4}$ K$^{-1}$ [25]. Considering the experimental sensitivity of the wavelength shift versus temperature change of methanol treated silk (*Silk II*), we can get the thermal-optic coefficient $dn_{eff}/(n_{eff}dT) = 2.4 \times 10^{-4}$ K$^{-1}$ in 1430 nm wavelength band, which means the silk fibroin toroid based sensor is dominated by thermal expansion.

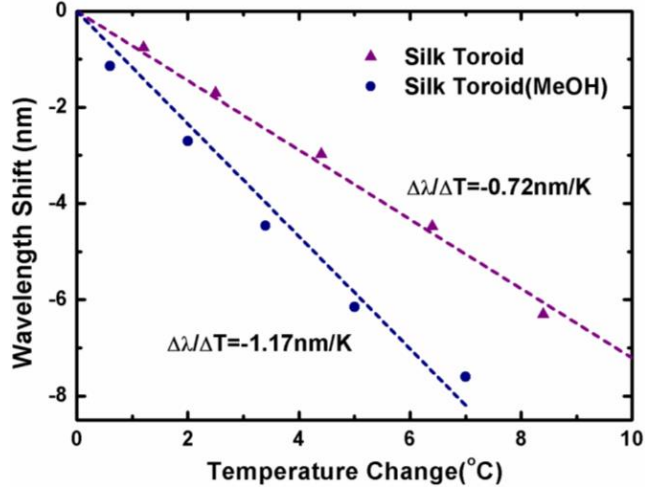

Fig. 4. Resonant wavelength shift versus the temperature change for silk microtoroids with (circle) and without (triangle) methanol (MeOH) treatment. Silk microtoroid donminated by *Silk I* displayed a sensitity of 0.67 nm/K, while methanol treated silk one showed a higher sensitivity of 1.17 nm/K.

## 5. Conclusions

In summary, we have demonstrated silk fibroin microtoroids on a flexible chip with $Q$ factors as high as $10^5$, limited by the absorption loss. A thermal sensing experiment was also realized in this silk fibroin microtoroid which exhibited a measured sensitivity of 1.17 nm/K, which is 8 times higher than the previous WGM resonator based thermal sensors thanks to the large thermal expansion coefficient of silk. This opens promising avenues to fabricate biodegradable and biocompatible protein-based microresonators on a flexible chip for biophotonics applications that use natural materials as their constituents, adding utility to the range of technical applications that are enabled by the use of structural proteins such as silk.

## Funding

National Science Foundation (NSF) (1264997); Office of Naval Research (N00014-13-1-0596).

## Acknowledgments

Linhua Xu thank Carlo Herbosa, Steven He Huang, Weijian Chen, Huzeyfe Yilmaz, Bo Peng, Faraz Monifi, Jiangang Zhu, Sahin Kaya Ozdemir, Keng-Ku Liu, and Srikanth Singamaneni for helpful discussions.